\documentclass[reprint,amsmath,amssymb,showkeywords, aps,nofootinbib]{revtex4-1}
\usepackage{graphicx}
\usepackage{dcolumn}
\usepackage{bm}

\begin{document}
\preprint{APS/123-QED}
\title{On Cosmic Strings, Infinite Line-Masses and Conformal Invariance}
\author{Reinoud Jan Slagter}
\altaffiliation[Also at ]{Physics Department, Amsterdam University}
\email{info@asfyon.com}
\affiliation{  ASFYON, Astronomisch Fysisch Onderzoek Nederland, The Netherlands}
\date{\today}
\begin{abstract}
We investigate the  conformal invariant Lagrangian of the self-gravitating U(1) scalar-gauge field and  find new features of the model on  the time-dependent axially symmetric Bondi-Marder spacetime.
By considering the conformal symmetry as exact at the level of the Lagrangian and broken in the vacuum, a consistent model is found with an exact solution of the vacuum Bondi-Marder spacetime $g_{\mu\nu}=\omega^2 \bar g_{\mu\nu}$, where $\omega$ is the conformal factor and $\bar g_{\mu\nu}$ the `un-physical` spacetime. If we try to match this vacuum solution onto the interior vortex solution of the coupled Einstein-scalar-gauge field, we need, besides the matching conditions, constraint equations in order to obtain a topological regular  description of the small-scale behaviour of the model. Probably, one needs the five-dimensional warped counterpart model, where the 5D dilaton field acts as a warp factor. Moreover, the tracelessness of the energy-momentum tensor could then be maintained by a contribution from the bulk.
\begin{description}
\item[PACS numbers]
04.50.-h; 04.30.Db; 04.20.Ex; 11.10.Lm; 98.80.Cq; 11.27.+d; 98.54.Aj; \\ 04.30.Tv; 11.10.Kk
\item[keywords]
cylindrical symmetry; infinite line-mass; cosmic strings; U(1) scalar-gauge field; \\ gravitational waves; conformal invariance
\end{description}
\end{abstract}
\keywords{cylindrical symmetry, infinite line mass, cosmic strings, U(1) scalar-gauge field, gravitational waves, conformal invariance}
\maketitle
\section{\label{sec:level1}Introduction}
The general relativity theory (GRT) admits stationary and static axially symmetric solutions of compact objects. The most famous one is the Kerr black hole solution. One believes that in the center of galaxies there is a rotating super-massive Kerr black hole. The Kerr solution is a member of the family of axially symmetric solutions of the Einstein equations. A legitimate question is then: are there other axially or cylindrically symmetric asymptotically flat solutions in GRT which can be connected to real physical objects. In the early days of the development of the GRT, many researchers tried to find other axially symmetric solutions with the correct asymptotic form. For an overview of these solutions, we refer to Stephani\cite{Step}, et al. and Islam\cite{Islam}. In order to understand these  axially symmetric objects in there full exposure, an interior solution is necessary with the correct matching conditions. It turns out, that this is a hard task, even for the Kerr solution.
It came as a big surprise that a quantum field on the righthand side of the equations of Einstein could lead to physically acceptable solutions. The most famous one is the coupled Einstein Abelian-Higgs model. The gauged  scalar field with the "Mexican-hat"-potential has lived up to its reputations. It was successful in the explanation of superconductivity, in the standard model of particle physics with the spontaneously broken symmetry by the  Brout-Englert-Higgs (BEH) mechanism and in the general relativistic solution of the self-gravitating Nielsen-Olesen(NO)-vortex\cite{Fel,Nielsen,Garf}. The NO-vortex has a cylindrical symmetry, so it would be quite natural to study, in this context, axially (or cylindrical) symmetric spacetimes. Further, it was realized that string-like field configurations could be produced within the framework of superstring theory. These so-called cosmic superstrings could play an interesting role in warped brane-world models\cite{Shir,Slag1,Slag2}
There is an additional motivation for studying axially symmetry: whereas a spherical mass surrounded by empty space is truly isolated, a cylindrical mass distribution will cause energy flow to and from infinity. It is believed that in the early stages of the universe, the  emission of gravitational waves will play an important role. So we will study here  radiating non-static models.
The study of cylindrical symmetric gravitational waves goes back a long way. Einstein and Rosen (ER)\cite{ER} showed that the free-field equations admit exact solutions corresponding to cylindrical gravitational waves.
Standard, one applies a pulse wave superimposed on the static solution and calulates the change in the metric components. Here we will apply the full coupled time-dependent equations.

Finally, there is another  tantalizing  argument for studying these vortex solution in gravity models, i.e., in context of conformal invariance. The Einstein-Hilbert action of gravity coupled to quantum fields can be reformulated by focusing on local conformal symmetry as an exact but spontaneously broken symmetry\cite{thooft1,thooft2,thooft3,thooft4,thooft5}.
This phenomenon could be used to tackle the  huge problem how to handle  the small-distance structure of this coupled system of a quantum field and gravity. Small time and distance scales seems not to be related to large time and distance scales.
In quantum field theory one usually works on Minkowski spacetime. However, curved spacetime will inevitable enter the field equations on small scales and gravity cannot be ignored.
The first task is then to construct a Lagrangian, where spacetime and the fields defined on it, are topological regular. This can be done by considering the scale factor( or warp factor in higher-dimensional models) as a
dilaton field besides the conformally coupled  scalar field.
The same procedure can be applied to the 4D spacetime and one can try to generate from (Ricci)-flat spacetimes physical acceptable spacetimes in the non-vacuum case.
It is known since the 70s\cite{thooft0}, that quantum field theory combined with Einstein's gravity  runs into serious problems. The Einstein-Hilbert(EH) action is non-renormalizable and it gives rise to intractable divergences at loop levels.
On very small scales, due to quantum corrections to GR, one must modify Einstein's gravity by adding higher order terms in the Lagrangian like  $R^2, R_{\mu\nu}R^{\mu\nu}$ or $R_{\mu\nu\sigma\tau}R^{\mu\nu\sigma\tau}$ (or combinations of them).
However, serious difficulties arise in these higher-derivative models, for example,  the occurrence of massive ghosts which cause unitary problems.
A next step is then to disentangle the functional integral over the dilaton field from the ones over the metric fields and matter fields.
Moreover, it is desirable that all beta-functions of the matter lagrangian, in combination with the dilaton field, disappear in order to fix  all the coupling constants of the model.
Further, conformal invariance of the action with matter fields implies that the trace of the energy-momentum tensor is zero. A theory based on a classical "bare" action which is conformally invariant, will lose it in quantum
theory as a result of renormalization and the energy-momentum tensor acquires a non-vanishing trace ( trace anomaly).
We consider here the breaking of conformal invariance in conventional Einstein theory and will not enter into details of these quantum-gravity problems (see for example the text book of Parker and Toms\cite{Parker}).
It is conjectures that conformal symmetry is  exact at the level of the Lagrangian and only broken in the vacuum, just as the very weakly interacting sub-eV axion can break the CP invariance in QCD and the BEH mechanism in standard model of particle physics. This approach can even be an alternative to supersymmetry and  the dark energy problem\cite{Mann1,Mann2}.
Because our axially symmetric model can easily transformed to spherical symmetry, it is clear that our conformal invariant study of the self-gravitating coupled scalar-gauge field on an axially symmetric spacetime
make sense  in studying the small scale properties.
In section 2 we will summarize the classical axially symmetric spacetimes that are of importance for our study.
In section 3 we formulate the self-gravitating abelian Higgs model on the Bondi-Marder spacetime.
In section 4  we reformulate the model in a conformal invariant setting and in section 5 we consider the interior vortex solution of the coupled Einstein-scalar-gauge field and compare the results with the 5D counterpart model.

\section{\label{sec:level2}The Classical Axially Symmetric Spacetime}
A major investigation on cylindrical and axial symmetric models  were done by Stachel\cite{Stachel}, Marder\cite{Marder} and Bondi\cite{Bondi}. The general class of vacuum cylindrical symmetric spacetimes can be written as
\begin{equation}
ds^2=e^{-2\pmb{\psi}}\Bigl[e^{2\pmb{\gamma}}(d\rho^2-dt^2)+\rho^2 d\varphi^2\Bigr]+e^{2\pmb{\psi}} d\varphi^2,\label{eq2-1}
\end{equation}
where $\pmb{\gamma}(t,\rho)$ and $\pmb{\psi}(t,\rho)$ represent two degrees of freedom of the gravitational field\footnote{{\tiny We use the boldface  Greek symbols for the metric components in order to make a distinction between the numerous constants entering the exact solutions and the field variables. Moreover, we will use the same symbols as used by Marder in his original paper.}}.
The scalar cylindrical wave equation for $\pmb{\psi}$ is decoupled from $\pmb{\gamma}$,
\begin{equation}
\pmb{\psi}_{tt}=\pmb{\psi}_{\rho\rho}+\frac{1}{\rho}\pmb{\psi}_\rho, \quad \pmb{\gamma}_t=2\rho\pmb{\psi}_t\pmb{\psi}_\rho, \quad \pmb{\gamma}_\rho=\rho (\pmb{\psi}_t^2+\pmb{\psi}_\rho^2)\label{eq2-2},
\end{equation}
and many solutions can be found. However, many features of these solutions are related to the well-studied  static and stationary axially symmetric solutions of a line-mass distribution, if one performs the transformation $t\rightarrow iz, z\rightarrow it$. An example is the famous Weyl solution (Ricci flat semi-infinite line (SILM)-mass)
\begin{eqnarray}
ds^2&=&-\alpha\Bigl[\epsilon(z-z_1)+\sqrt{(z-z_1)^2+\rho^2}\Bigr]^{2c_1}dt^2 \cr
&+&\frac{1}{\alpha}\frac{\Bigl[\epsilon (z-z_1)+\sqrt{(z-z_1)^2+\rho^2}\Bigr]^{4c_1^2-2c_1}}{\Bigl[2\sqrt{(z-z_1)^2+\rho^2}\Bigr]^{4c_1^2}}(ds^2+dz^2)\cr
&+&\frac{\rho^2}{\alpha}\Bigl[\epsilon(z-z_1)+\sqrt{(z-z_1)^2+\rho^2}\Bigr]^{-2c_1}d\varphi^2.\label{eq2-3}
\end{eqnarray}
For $\epsilon =-1$ the SILM extends to $+\infty$ and $z_1$ is the position of the beginning of the infinite line-mass. $c_1$ is related to the mass of the SILM and $\alpha$ to the angle deficit. See also Appendix A.
This metric leads to the Schwarzschild metric by a suitable transformation.
In the context of gravitational wave emission, Marder found that when gravitational waves are emitted for a short time, there must be an interaction with the mass distribution of the "cylinder" and one needs two parameters for the description of the exterior wave solution.
For  the Schwarzchild exterior, one needs only one. The crucial difference is that whereas a spherical mass surrounded by empty space is truly isolated, a cylindrical mass distribution will cause energy flow to and from infinity.  If an initially static solution emits a pulse of radiation, then there will be a change in  the value of one of the two parameters. In other to handle this these complexities, Thorne\cite{Thorne} introduced the notion of C-energy to describe the transport of cylindrical gravitational waves.
\section{\label{sec:level3}Cosmic String Solution on the Bondi-Marder Cylindrical Symmetric Spacetime}
\subsection{\label{sec:level3-1}The Field Equations of the Model}
Let us consider the cylindrical symmetric spacetime suitable for the description of matter-filled regions of space.
\begin{equation}
ds^2=e^{-2\pmb{\psi}}\Bigl[e^{2\pmb{\gamma}}(d\rho^2-dt^2)+\rho^2d\varphi^2\Bigr]+e^{2\pmb{\psi}+2\pmb{\mu}}dz^2,\label{eq3-1}
\end{equation}
where $\pmb{\psi}, \pmb{\gamma}$ and $\pmb{\mu}$ are functions of $t$ and $\rho$. The spacetime Eq.(\ref{eq2-1}) is not suitable, because we would have $T_{tt}+T_{\rho\rho}$=0. This is not the case for cosmic strings.
Further, we don't make yet a gauge choice, such as the usual one $e^{\pmb{\mu}}=\rho$.
For the matter distribution we choose the scalar-gauge field
\begin{equation}
\Phi=\eta X e^{in\varphi},\qquad A_\mu=[0,0,0,\frac{P-n}{e}]\label{eq3-2}
\end{equation}
with potential $V=\frac{1}{8}\lambda(\Phi \Phi^*-\eta^2)^2$. X and P are functions of $t$ and $\rho$. The vacuum expectation value of the Higgs field is $\eta$, n is the winding number, $\lambda$ the Higgs coupling constant and e the electric charge of a Cooper pair.
For a detailed treatment of this $\Phi^4$ gauge model, which is heuristically equivalent with the Ginsburg-Landau theory of superconductivity,  we refer to Felsager\cite{Fel} and Vilenkin\cite{Vil}.
The starting point is the action
\begin{eqnarray}
{\cal S}=\int d^4x\sqrt{- g}\Bigl\{\frac{1}{2\kappa^2}\Bigl(R-2\Lambda)-\frac{1}{2} D_\alpha\Phi( D^\alpha\Phi)^* \cr
-\frac{1}{4}F_{\alpha\beta}F^{\alpha\beta}-\frac{1}{8}\lambda(\Phi^2-\eta^2)^2\Bigr\},\label{eq3-3}
\end{eqnarray}
with $F_{ab}=\partial_a A_b-\partial_b A_a$ the electro-magnetic field tensor. The gauge-covariant derivative is $D_a=\partial_a+ieA_a$.
The field equations become ( for the moment with $\Lambda =0$)
\begin{equation}
G_{\mu\nu}=\kappa^2 T_{\mu\nu},\label{eq3-4}
\end{equation}
\begin{equation}
D_\mu D^\mu \Phi -2\frac{\partial V}{\partial \Phi^*}=0,\label{eq3-5}
\end{equation}
and
\begin{equation}
\nabla^\nu F_{\mu\nu}-\frac{1}{2}ie\Bigl[\Phi(D_\mu\Phi)^*-\Phi^*(D_\mu \Phi)\Bigr] = 0,\label{eq3-6}
\end{equation}
with $T_{\mu\nu}$ the energy momentum tensor. Written out in components, the equations become
\begin{eqnarray}
\partial_{tt}\pmb{\psi}&=&\partial_{\rho\rho}\pmb{\psi}+\frac{1}{\rho}\partial_\rho\pmb{\psi}+\partial_\rho\pmb{\psi}\partial_\rho\pmb{\mu}-\partial_t\pmb{\psi}\partial_t\pmb{\mu}-\frac{1}{\rho}\partial_\rho\pmb{\mu} \cr
&+&\kappa^2\Bigl[e^{2\pmb{\psi}}\frac{\partial_t P^2-\partial_\rho P^2}{2r^2e^2}-\frac{1}{8}\lambda\eta^4e^{2\pmb{\gamma}-2\pmb{\psi}}(X^2-1)^2 \cr
&-&\eta^2e^{2\pmb{\gamma}}\frac{X^2P^2}{\rho^2} \Bigr],\label{eq3-7}
\end{eqnarray}
\begin{eqnarray}
\partial_{tt}&\pmb{\mu}&=\partial_{\rho\rho}\pmb{\mu}+\frac{2}{\rho}\partial_\rho\pmb{\mu}+\partial_\rho\pmb{\mu}^2-\partial_t\pmb{\mu}^2 \cr
&+&\kappa^2\Bigl[\frac{1}{4}\lambda\eta^4e^{2\pmb{\gamma}-2\pmb{\psi}}(X^2-1)^2
+\eta^2e^{2\pmb{\gamma}}\frac{X^2P^2}{\rho^2}\Bigr],\label{eq3-8}
\end{eqnarray}
\begin{eqnarray}
\partial_{tt}\pmb{\gamma}&=&\partial_{\rho\rho}\pmb{\gamma}+\partial_\rho\pmb{\psi}^2-\partial_t\pmb{\psi}^2+2(\partial_\rho\pmb{\psi}\partial_\rho\pmb{\mu}-\partial_t\pmb{\psi}\partial_t\pmb{\mu}) \cr
-\frac{2}{\rho}\partial_\rho\pmb{\mu}
&+&\frac{\kappa^2}{2}\Bigl[e^{2\pmb{\psi}}\frac{\partial_t P^2-\partial_\rho P^2}{2r^2e^2}+\eta^2(\partial_\rho X^2-\partial_t X^2)\cr
&-&\frac{1}{4}\lambda\eta^4e^{2\pmb{\gamma}-2\pmb{\psi}}(X^2-1)^2-3\eta^2e^{2\pmb{\gamma}}\frac{X^2P^2}{\rho^2}\Bigr],\label{eq3-9}
\end{eqnarray}
\begin{eqnarray}
\partial_{tt}X=\partial_{\rho\rho}X+\frac{\partial_\rho X}{\rho}+\partial_\rho X\partial_\rho\pmb{\mu}-\partial_t X\partial_t\pmb{\mu} \cr
-e^{2\pmb{\gamma}}\frac{XP^2}{\rho^2}-\lambda\eta^2e^{2\pmb{\gamma} -2\pmb{\psi}}X(X^2-1),\label{eq3-10}
\end{eqnarray}
\begin{eqnarray}
\partial_{tt}P=\partial_{\rho\rho}P-\frac{\partial_\rho P}{\rho}+2(\partial_\rho P\partial_\rho\pmb{\psi}-\partial_t P\partial_t\pmb{\psi}) \cr
+\partial_\rho P\partial_\rho\pmb{\mu}-\partial_t P\partial_t\pmb{\mu} +e^2\eta^2e^{2\pmb{\gamma}-2\pmb{\psi}}PX^2.\label{eq3-11}
\end{eqnarray}
These equation can be solved numerically and the well-know solutions of the scalar field and gauge field are expected\cite{Vil}. There is still a constraint equation from the $(t,r)$ component.
In Figure 1 we plotted a typical numerical solution. At the core of the string we applied the  boundary conditions of the gauge string solution\cite{Garf} and took for $\gamma$ and $X$ an initial disturbance.
\begin{figure}[h]
\centerline{
\includegraphics[scale=.20]{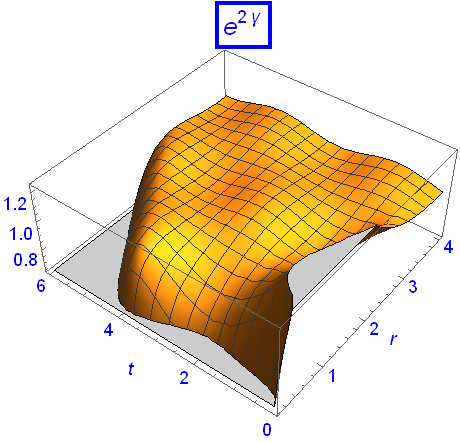}
\includegraphics[scale=.20]{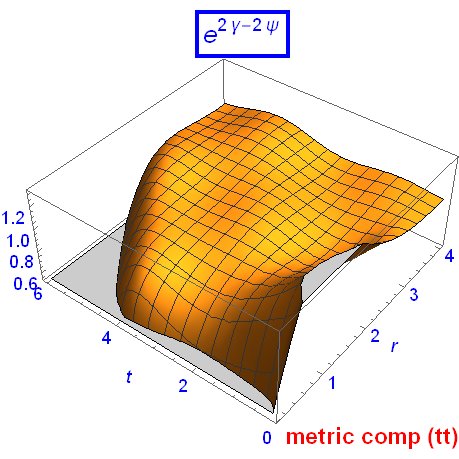}}
\centerline{
\includegraphics[scale=.20]{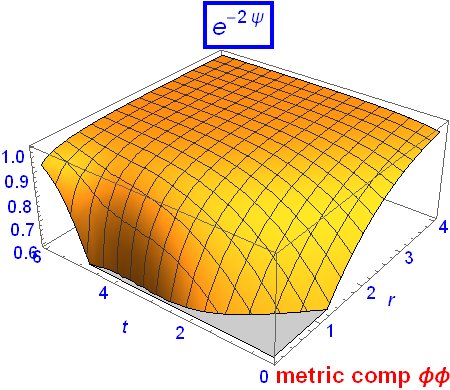}
\includegraphics[scale=.20]{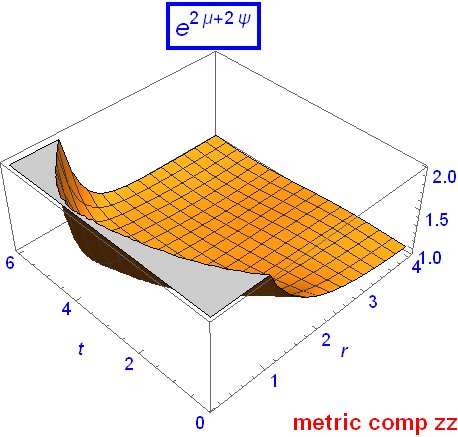}}
\centerline{
\includegraphics[scale=.20]{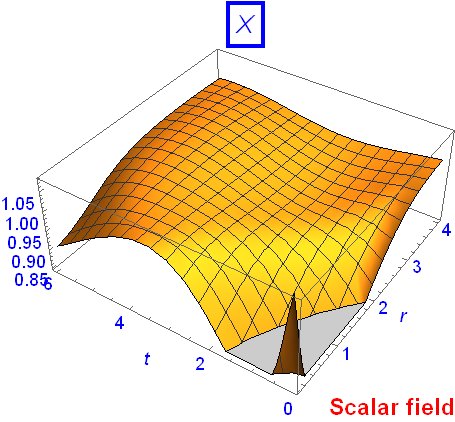}
\includegraphics[scale=.20]{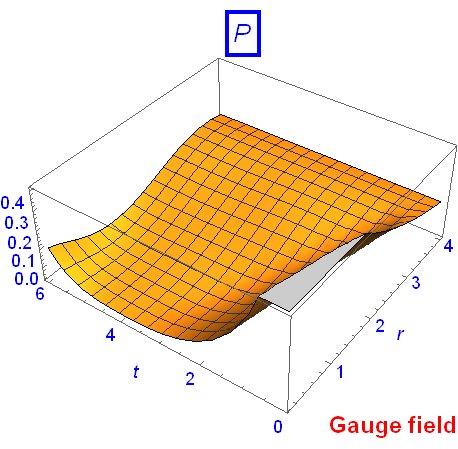}}
\caption{{\bf A typical solution of the interior non-vacuum model of Eq.(\ref{eq3-7})-Eq.(\ref{eq3-11}). }}
\label{fig1}
\end{figure}
The matching conditions could even deliver restrictions for the parameters of the scalar-gauge field, i.e., $\lambda$ and  $\eta$.
As  Marder\cite{Marder} already pointed out, from a physical point of view, the concept of a smooth pulse wave of finite duration ( apart from a residual "tail") is more acceptable than a wave motion that has to take an infinite time to its present state.
In order to match this solution with the exterior vacuum solution, we need the exact solution of the exterior. This is done in the next section.
\subsection{\label{sec:level3-2}The Exterior vacuum Solution}
A general exact solution of the exterior vacuum outside the matter distribution of Eq.(\ref{eq3-7}-\ref{eq3-11}) is
\begin{figure}[h]
\centerline{\includegraphics[scale=.22]{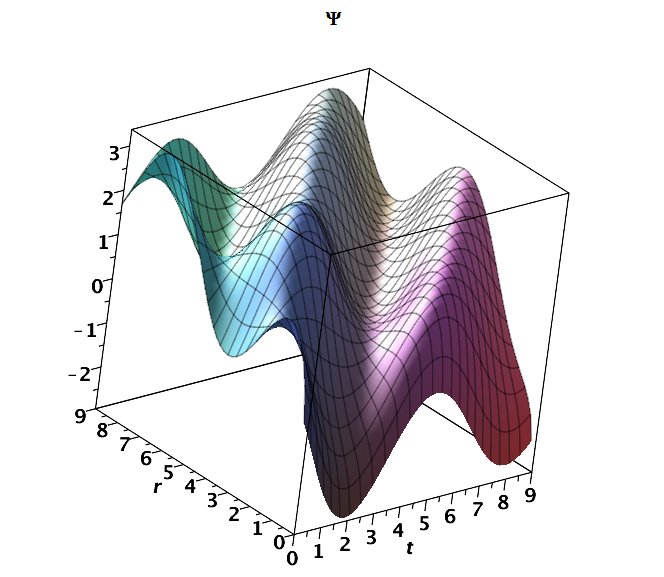}
\includegraphics[scale=.18]{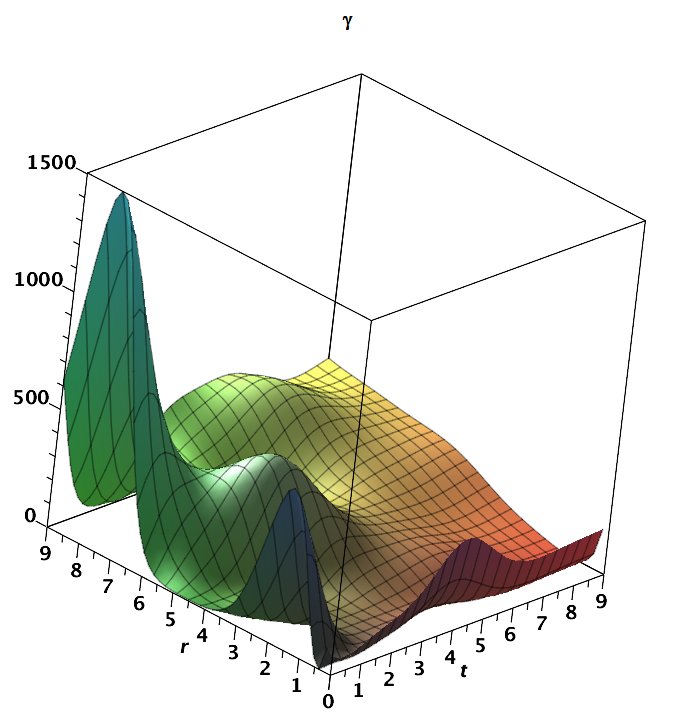}}
\centerline{
\includegraphics[scale=.23]{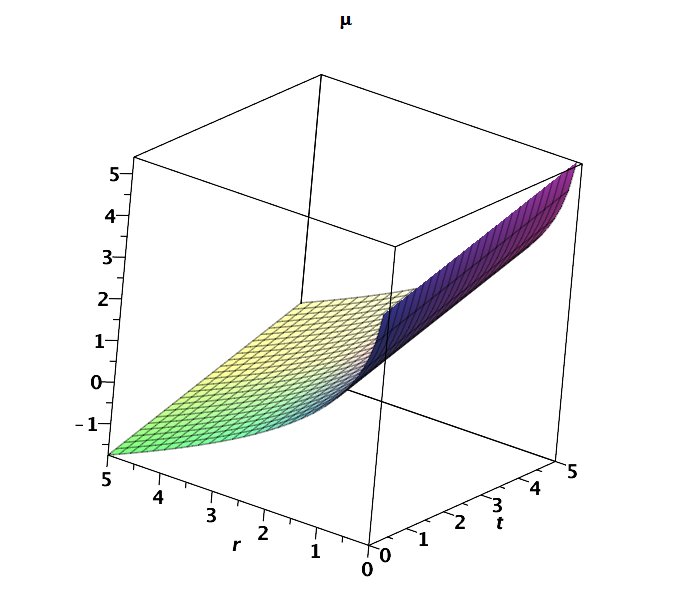}}
\caption{{\bf A typical solution of an exact solution of ${\pmb\psi}, {\pmb\gamma} $ and ${\pmb\mu}$ Eq.(\ref{eq3-12})-Eq.(\ref{eq3-14}) }}
\label{fig2}
\end{figure}
\begin{eqnarray}
\pmb{\mu}=\ln\Bigl[\frac{\beta_1\sinh(\sqrt{c_1}\rho)+\beta_2\cosh(\sqrt{c_1}\rho)}{\sqrt{c_1}\rho}\Bigr]\cr
+\ln\Bigl[\frac{\beta_3e^{2\sqrt{c_1}t}+\beta_4}{2\sqrt{c_1}}\Bigr]-\sqrt{c_1}t,\label{eq3-12}
\end{eqnarray}
\begin{equation}
\pmb{\psi}={\cal G}_1(\rho-t)+e^{-\sqrt{c_1}t}{\cal G}_2(\rho +t)+\ln(\rho),\label{eq3-13}
\end{equation}
\begin{eqnarray}
\pmb{\gamma} &=& {\cal G}_3(\rho -t)+{\cal G}_4(\rho+t)+\ln(\rho)-\frac{1}{4}e^{-2\sqrt{c_1}t}\sin^2(\rho +t) \cr
&+&2e^{-\sqrt{c_1}t}\Bigl[\sin(\rho +t)+\frac{1}{c_1+4}(\cos(2t)\cr
&-&\cos(2\rho)+\sqrt{c_1}\sin(2\rho))\Bigr],\label{eq3-14}
\end{eqnarray}
with ${\cal G}_i $ arbitrary functions in the arguments and $\beta_i , c_1$ some constants. We took, for the time being,  $\beta_1=\beta_2=\beta_3=1, \beta_4=0$ and ${\cal G}_1=\sin(\rho -t)$ and ${\cal G}_2=\cos(\rho +t)$.
In figure 2 we plotted this typical solution.

However, for our model, we will use another simpler exact solution
\begin{eqnarray}
ds^2&=&\frac{1}{D}e^{G\rho+Ht}(t+A)^{2E(E+1)}(B\rho+C)^{2\frac{F}{B}(\frac{F}{B}-1)}\cr
&\times & [-dt^2+d\rho^2]+D (t+A)^{2(E+1)}(B\rho+C)^{\frac{2F}{B}}dz^2\cr
&+&\frac{1}{D}(t+A)^{-2E}(B\rho+C)^{2(1-\frac{F}{B})}d\varphi^2, \label{eq3-15}
\end{eqnarray}
where we have 8 constants. So we can impose boundary conditions at the core of the string which will contain the parameters of the scalar-gauge field, such as the vacuum expectation value and gauge coupling constant.
This metric will be Ricci flat for $F=B(E+2)$ or $F=EB$  and suitable values for  G,H and K.
There are some other possible solutions, which we will not consider here.

There is an interesting relation with the spacetime of the SILM and ILM, already mentioned in section 2 for the spacetime of Eq.(\ref{eq2-1}).
In the next section we will consider this relation ( see also Appendix A).
\subsection{\label{sec:level 3-3}Connection with Accelerated Observers and the (Semi-)Infinite Line-mass}
The features of an infinite line-mass and semi-infinite line-mass was investigated  decades ago by many physicists\cite{Marder,Bonner1,Bonner2,Bonner3,Bonner4,Bondi}.
In the Appendix A we summarized the features of the (semi-)-infinite line mass.
Let us start with two most simplified flat cases of Eq.(\ref{eq3-15})
\begin{equation}
ds^2=-dt^2+d\rho^2+(t+A)^2dz^2+(B\rho+C)^2d\varphi^2,\label{eq3-18}
\end{equation}
and
\begin{equation}
ds^2=-dt^2+d\rho^2+(B\rho+C)^2dz^2+(t+A)^2d\varphi^2.\label{eq3-19}
\end{equation}
The first one  can be transformed to Minkowski chart (A=0, B=1 and C=0)
\begin{equation}
ds^2=-dT^2+d\rho^2+dZ^2+\rho^2d\varphi^2\label{eq3-20}
\end{equation}
by
\begin{eqnarray}
t=\sqrt{T^2-Z^2},\qquad z=arctanh\Bigl(\frac{Z}{T}.\Bigr)\label{eq3-21}
\end{eqnarray}
\begin{figure}[h]
\centerline{\includegraphics[scale=.195]{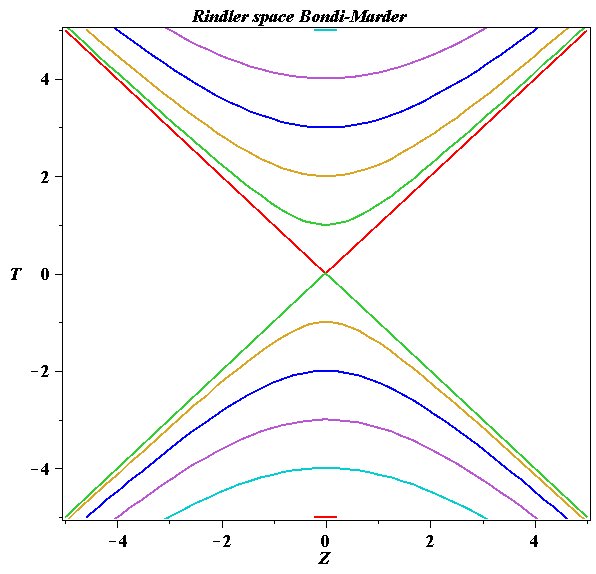}
\includegraphics[scale=.22]{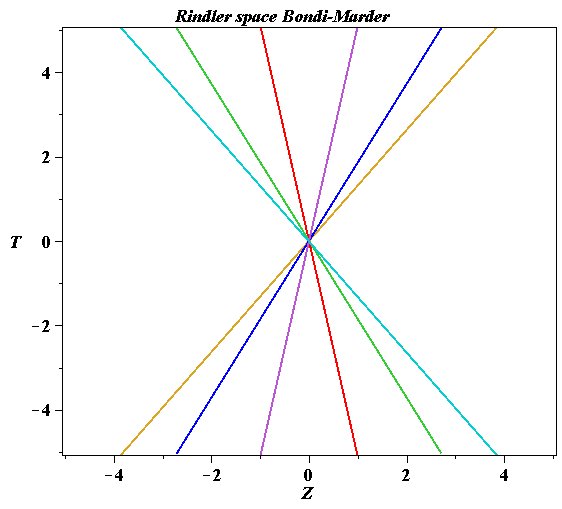}}
\vspace{10mm}
\caption{{\bf Rindler diagram of the spacetime Eq.(\ref{eq3-18}). Left: plot of $T=\sqrt{t^2+Z^2}$ for $t$=-6...6.
Right: plot of $T=Z/\tanh(z)$ for $z$=-0.4...0.4. See Eq.(\ref{eq3-21}).  }}
\label{fig3}
\end{figure}
In figure 3 we plotted the Rindler diagram.
So without any reference to the "rod"-features of the solution, we obtain the typical Rindler wedge of the SILM of Eq.(\ref{eqA-5}) of the  Appendix A, if we perform first the usual transformation
$t\rightarrow iz, z\rightarrow it$ from $(\rho, z)$ dependency to $(\rho, t)$ dependency  followed by  a Wick rotation $z\rightarrow iz$.
If one considers the coupled Einstein-scalar-gauge field (see section 5), one expect the Unruh effect for accelerated observers\cite{Unruh}. These observers will detect a black body spectrum.
It is worth to investigate this issue in our model.

The transition to the non-vacuum situations causes problems with the interpretation of a line mass. The dimension of the core of the string cannot be taken infinite thin (wire-approximation. In  Appendix B we summarized these problems.
\section{\label{sec:level 4}Conformal Invariance}
Performing a conformal transformation on a spacetime manifold, means that we change our standard measuring rods and clocks. This change is not the same in different points in spacetime. In other words, we multiply the spacetime metric by a kind of scalar field (or dilaton field).
The notion of conformal invariance in GR is properly handled in the text books of Parker, et al.\cite{Parker} and Wald\cite{Wald}. In Appendix C we briefly summarized some important features.
Let us rewrite the metric of Eq.(\ref{eq3-1}) as
\begin{equation}
ds^2=\pmb{\omega}^2\Bigl[-dt^2+d\rho^2+e^{2\pmb{\tau}}dz^2+\rho^2 e^{-2\pmb{\gamma}}d\varphi^2\Bigr],\label{eq4-1}
\end{equation}
where we wrote $e^{4\pmb{\psi}+2\pmb{\mu}-2\pmb{\gamma}}\equiv e^{2\pmb{\tau}}$ and $\pmb{\omega}^2\equiv e^{2\pmb{\gamma} -2\pmb{\psi}}$. So we define a metric $ \bar g_{\mu\nu}$ by $g_{\mu\nu}=\pmb{\omega}^2\bar g_{\mu\nu}$.
From the Einstein equations for the metric $\bar g_{\mu\nu}$  we obtain the PDE's
\begin{equation}
\partial_{tt}\pmb{\bar\tau}= \partial_{\rho\rho}\pmb{\bar\tau}+\partial_\rho\pmb{\bar\tau}^2-\partial_t\pmb{\bar\tau}^2,\label{eq4-2}
\end{equation}
\begin{equation}
\partial_{tt}\pmb{\bar\gamma}= \partial_{\rho\rho}\pmb{\bar\gamma}+\partial_t\pmb{\bar\gamma}^2-\partial_\rho\pmb{\bar\gamma}^2+\frac{2}{\rho}\partial_\rho\pmb{\bar\gamma}.\label{eq4-3}
\end{equation}
The solutions for $\pmb{\bar\tau}$ and $\pmb{\bar\gamma}$ of the $\bar g_{\mu\nu}$ are
\begin{eqnarray}
\pmb{\bar\tau}&=&{\cal F}(t+\rho),\cr
\pmb{\bar\gamma}&=&-\ln\Bigl[\frac{(\epsilon_1 e^{2\sqrt{c1}t}+\epsilon_2)(\epsilon_3\sinh(\sqrt{c1}r)+\epsilon_4\cosh(\sqrt{c1}r))}{2c_1r}\Bigr] \cr
&+&\sqrt{c_1}t,\label{eq4-4}
\end{eqnarray}
with ${\cal F}(t+\rho)$ an arbitrary function of $(t+\rho)$ and $\epsilon_i$ constants.  Further, there is a constraint equation in first order derivatives of $\pmb{\bar\tau}$ and $\pmb{\bar\gamma}$.
In figure 4 we plotted an example of this solution, which is asymptotically Minkowski.
\begin{figure}[h]
\centerline{\includegraphics[scale=.21]{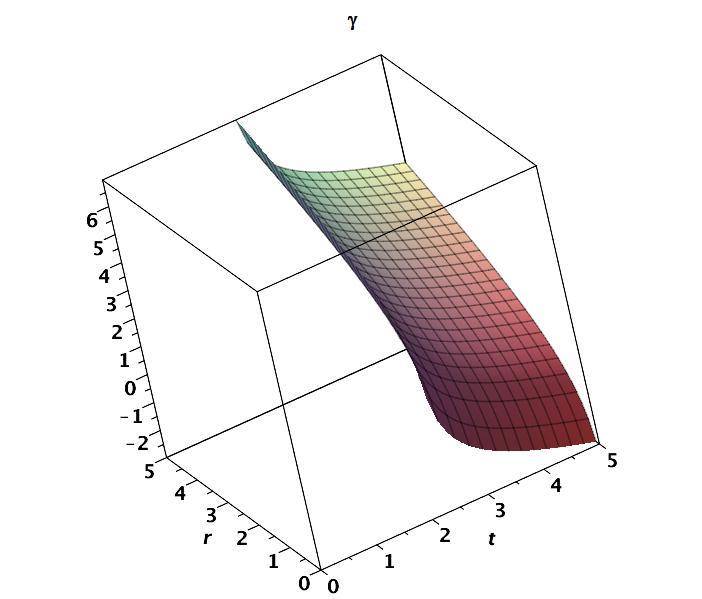}
\includegraphics[scale=.21]{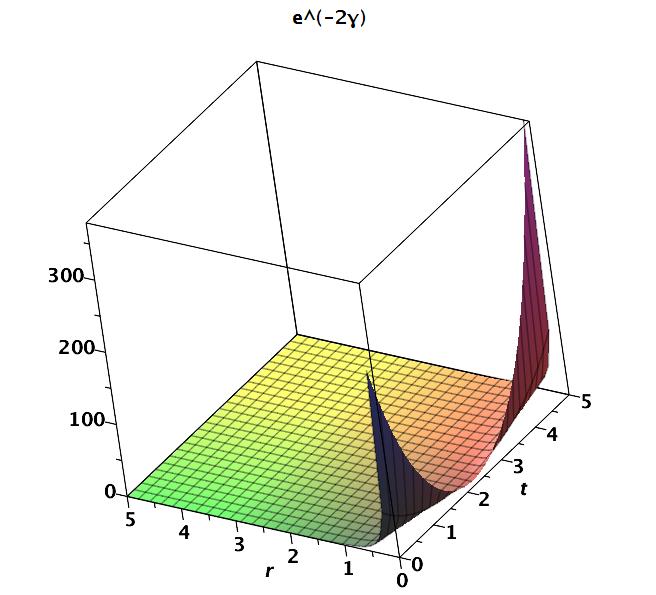}}
\caption{{\bf Exact solution for the metric component  $\bar g_{\varphi\phi}$ of Eq.(\ref{eq4-4})}}
\label{fig4}
\end{figure}

A special solution of $\bar g_{\mu\nu}$  for suitable values of the constants is
\begin{eqnarray}
\bar ds^2&=&-dt^2+d\rho^2+(\sigma_1 t +\sigma_2\rho +\sigma_3)^2dz^2 \cr
&+&(\sigma_4\rho+\sigma_5)^2(\sigma_6 t +\sigma_7)^2 d\varphi^2,\label{eq4-5}
\end{eqnarray}
with $\sigma_i$ constants. This metric  is equivalent to the flat spacetime of Eq.(\ref{eq3-18}) or Eq.(\ref{eq3-19}) for suitable $\sigma_i$, i.e., a uniformly accelerated metric in the $z$-direction along the SILM.
So we can say that our $\bar g_{\mu\nu}$ is a "scaled" version of the SILM.
\subsection{\label{sec:level 4-2} The conformal invariant vacuum model}
We are not really interested in the solution for $\bar g_{\mu\nu}$ of the last section, but merely in a solution of the conformal invariant model.
If we substitute
\begin{equation}
g_{\mu\nu}=\pmb{\omega}^2\bar g_{\mu\nu}\label{eq4-6}
\end{equation}
into the EH action, then one easily verify that
\begin{eqnarray}
{\cal L}^{EH\pmb{\omega}}=\frac{\sqrt{- g}}{2\kappa^2}\Bigl(\pmb{\omega}^2 R+6 g^{\mu\nu}\partial_\mu\pmb{\omega}\partial_\nu\pmb{\omega}\Bigr)\label{eq4-7}
\end{eqnarray}
is conformally invariant by $\omega \rightarrow \frac{1}{\Omega}\omega$, $ g_{\mu\nu} \rightarrow \Omega^2 g_{\mu\nu}$ (see Appendix C).
Variation with respect to $ g_{\mu\nu}$ results in the Einstein equation
\begin{eqnarray}
\pmb{\omega}^2 G_{\mu\nu}&=&\nabla_\mu\nabla_\nu\pmb{\omega}^2- g_{\mu\nu}\nabla_\alpha\nabla^\alpha\pmb{\omega}^2-6\partial_\mu\pmb{\omega}\partial_\nu\pmb{\omega} \cr
&+&3 g_{\mu\nu}\partial_\alpha\pmb{\omega}\partial^\alpha\pmb{\omega}\equiv T_{\mu\nu}^{(\pmb{\omega})}.\label{eq4-8}
\end{eqnarray}
Variation of Eq.(\ref{eq4-7}) with respect to $\pmb{\omega}$ yields the well-known conformal invariant equation
\begin{equation}
\nabla^\mu\partial_\mu\pmb{\omega}-\frac{1}{6}\pmb{\omega}  R=0.\label{eq4-9}
\end{equation}
One can easily verify that {\bf TR}$[ G^{\mu\nu}-\frac{1}{\pmb{\omega}^2} T_{\mu\nu}^{(\pmb{\omega})}]=0$. So the trace of any matter field contribution must be zero.
From the Einstein equations, Eq.(\ref{eq4-8}) and  the $\pmb{\omega}$ equation,  Eq.(\ref{eq4-9}), we can separate  the equation for $\pmb{\omega}$
\begin{equation}
\partial_{tt}\pmb{\omega}=\partial_{\rho\rho}\pmb{\omega}+\frac{1}{\pmb{\omega}}(\partial_t\pmb{\omega}^2-\partial_r\pmb{\omega}^2)+\upsilon\pmb{\omega}.\label{eq4-10}
\end{equation}
The solution is
\begin{equation}
\pmb{\omega}={\cal A}e^{\frac{1}{2}\zeta_1(\rho^2+t^2)-\frac{1}{2}\upsilon\rho^2+\zeta_2 t+ r},\label{eq4-11}
\end{equation}
with $\zeta_i$  and ${\cal A}$
integration constants. The constant $\upsilon$ enters the equation by the separation of variables in the Einstein equations.
A typical example is plotted in figure 5.
\begin{figure}[h]
\centerline{
\includegraphics[scale=.2]{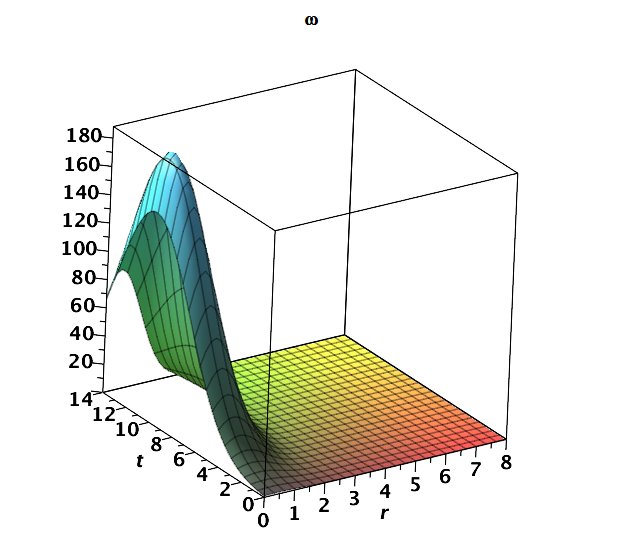}
\includegraphics[scale=.2]{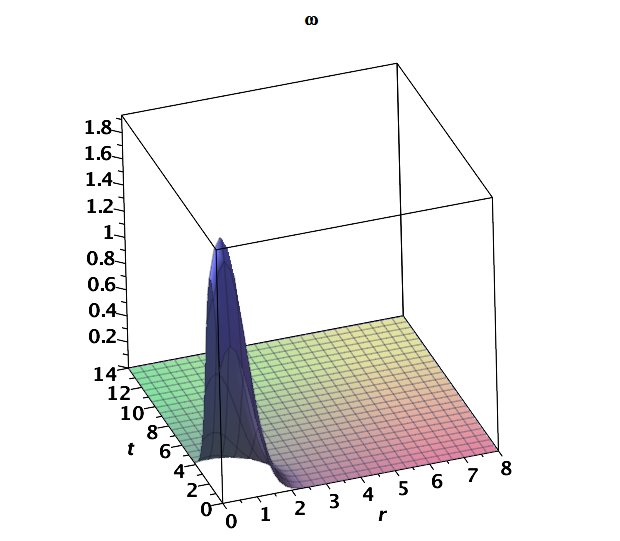}}
\centerline{
\includegraphics[scale=.2]{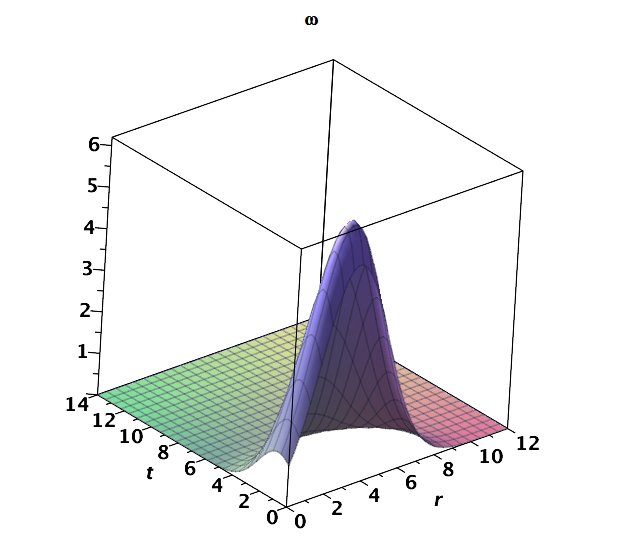}
\includegraphics[scale=.2]{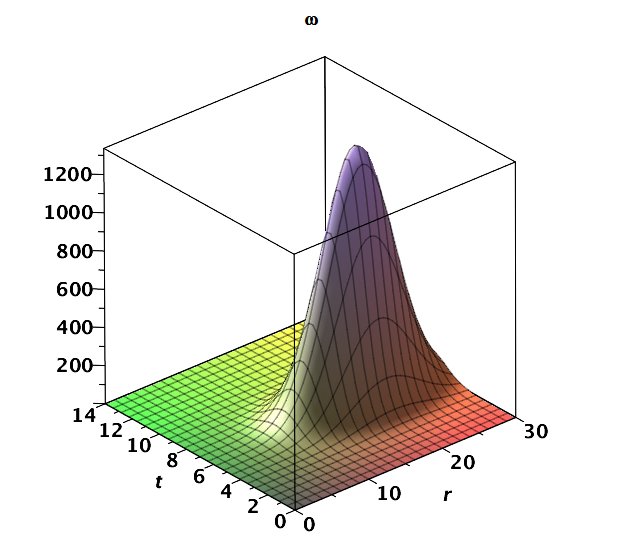}}
\caption{{\bf Some example plots of  $\pmb{\omega}$ of Eq.(\ref{eq4-11}) for different values of the constants $c_i$ and $\upsilon$. }}
\label{fig5}
\end{figure}
On very small scales for $t\rightarrow 0$, no singular behavior is  observed and   $\pmb{\omega}$ approaches a "scale" factor ${\cal A}$, which can be set as small as possible, determined by boundary conditions.
The equations for $\pmb{\tau}$ and $\pmb{\gamma}$ become (compare with Eq.(\ref{eq4-2}), Eq.(\ref{eq4-3}))
\begin{eqnarray}
\partial_{tt}\pmb{\tau}&=&\partial_{\rho\rho}\pmb{\tau}+\partial_\rho\pmb{\tau}^2-\partial_t\pmb{\tau}^2+\frac{2}{\pmb{\omega}}(\partial_\rho\pmb{\omega}\partial_\rho\pmb{\tau}-\partial_t\pmb{\omega}\partial_t\pmb{\tau})\cr
&+&\frac{1}{\pmb{\omega}^2}(\partial_\rho\pmb{\omega}^2-\partial_t\pmb{\omega}^2)\label{eq4-12}
\end{eqnarray}
and
\begin{eqnarray}
\partial_{tt}\pmb{\gamma}&=&\partial_{\rho\rho}\pmb{\gamma}+\partial_t\pmb{\gamma}^2-\partial_\rho\pmb{\gamma}^2
+\frac{2}{\pmb{\omega}}(\partial_\rho\pmb{\omega}\partial_\rho\pmb{\gamma}-\partial_t\pmb{\omega}\partial_t\pmb{\gamma})\cr
&+&\frac{2}{\rho}\partial_\rho\pmb{\gamma}+\frac{1}{\pmb{\omega}^2}(\partial_t\pmb{\omega}^2-\partial_\rho\pmb{\omega}^2)-\frac{2}{\rho\pmb{\omega}}\partial_\rho\pmb{\omega}.\label{eq4-13}
\end{eqnarray}
Together with the solution for $\pmb{\omega} $ of Eq.(\ref{eq4-11}), we  can solve these equations exactly ( for some choices of the parameters). The resulting metric becomes
\begin{eqnarray}
ds^2 = {\cal A}^2\Bigl\{e^{\zeta_1(\rho^2+t^2)-\upsilon\rho^2+2\zeta_2t+2\rho}(-dt^2+d\rho^2)\cr
+\Bigl[\frac{(e^{\sqrt{\vartheta_1}t}+\delta_1 e^{-\sqrt{\vartheta_1}t})^2 (e^{\sqrt{\vartheta_2}\rho}+\delta_2 e^{-\sqrt{\vartheta_2}\rho})^2}{\vartheta_1\vartheta_2}\Bigr]dz^2 \cr
+\Bigl[\frac{(e^{\sqrt{\vartheta_1^*}t}+\delta_3 e^{-\sqrt{\vartheta_1^*}t})^2 (e^{\sqrt{\vartheta_2^*}\rho}+\delta_4 e^{-\sqrt{\vartheta_2^*}\rho})^2}{\vartheta_1^*\vartheta_2^*}\Bigr]\Bigr\},\label{eq4-14}
\end{eqnarray}
where $\vartheta_1\equiv(\zeta_2^2+c_1+\zeta_1), \vartheta_2\equiv(\zeta_2^2+\upsilon+c_1+\zeta_1), \vartheta_1^*\equiv(\zeta_2^2-c_2+\zeta_1), \vartheta_2^*\equiv(\zeta_2^2+\upsilon-c_2+\zeta_2)$. $c_i$ are integration constants.
One cann easily verify that the Bianchi identities are fulfilled. A constraint equation is found for $\partial_t\omega$.
At the boundary $\rho=\rho_0$ we have now  exact values of the $\pmb{\gamma}, \pmb{\tau}$ and $\pmb{\omega}$, which can be used for the interior solution.
For  $\upsilon=0$ and $c_1=-c_2$, the Ricci-scalar and Weyl tensor are zero. Constraint equations can be found for Ricci-flat solutions.
Summarized, we can now generated  exact ($\Omega$-)conformal equivalent spacetimes Eq.(\ref{eq4-14}) from $g_{\mu\nu}$, which can be (Ricci) flat spacetimes.
\section{\label{sec:level 5}Breaking of the conformal invariance}
\subsection{\label{sec:level 5-1}The non-vacuum interior}
Now we extend the massless solution of section 4 to the matter filled region of the scalar gauge field of section 3.
This will break the conformal invariance, because the massive scalar-gauge field breaks the tracelessness of the energy momentum tensor.
Since the coordinates $\rho$ and $t$  for the interior have different scalings, we first write the interior spacetime of Eq.(\ref{eq3-1}) as
\begin{equation}
ds^2=e^{-2\pmb{\psi}}\Bigl[e^{2\pmb{\gamma}}(dr^2-dT^2)+r^2d\varphi^2\Bigr]+e^{2\pmb{\psi}+2\pmb{\mu}}dz^2,\label{eq5-1}
\end{equation}
where have re-labeled the interior radial coordinate as $r$ and the time coordinate as $T$.
If we substitute
\begin{equation}
g_{\mu\nu} =\omega^2 \tilde g_{\mu\nu}, \qquad \Phi=\frac{1}{\omega}\tilde\Phi,\label{eq5-2}
\end{equation}
in  the Lagrangian, we obtain
\begin{eqnarray}
{\cal S}&=&\int d^4x\sqrt{- \tilde g}\Bigl\{-\frac{1}{12}\Bigl(\tilde\Phi\tilde\Phi^*+\bar\omega^2\Bigr) \tilde R \cr
&-&\frac{1}{2}\Bigl( D_\alpha\tilde\Phi( D^\alpha\Phi)^*+\partial_\alpha\bar\omega\partial^\alpha\bar\omega\Bigr)
-\frac{1}{4}F_{\alpha\beta}F^{\alpha\beta} \cr
&-&V(\tilde\Phi ,\bar\omega)-\frac{1}{36}\kappa^2\Lambda\bar\omega^4\Bigr\},\label{eq5-3}
\end{eqnarray}
which is local  invariant under the transformation $g_{\mu\nu}\rightarrow\Omega^2 g_{\mu\nu}, \Phi \rightarrow \frac{1}{\Omega}\Phi$ and $\omega\rightarrow \frac{1}{\Omega}\omega$.
We redefined $\bar\omega^2\equiv-\frac{6\omega^2}{\kappa^2}$.
Varying the Lagrangian, we obtain the field equations for the metric components, the scalar field, the gauge field and the "dilaton" field
\begin{eqnarray}
\tilde G_{\mu\nu}&=&\frac{1}{(\bar\omega^2 +\tilde\Phi\tilde\Phi^*)}\Bigl(T_{\mu\nu}^{(\bar\omega)}+T_{\mu\nu}^{(\tilde\Phi,c)}+\tilde T_{\mu\nu}^{(A)}+\frac{1}{6}\tilde g_{\mu\nu}\Lambda_{eff}\kappa_4^2\bar\omega^4 \cr
&+&\tilde g_{\mu\nu}V(\tilde\Phi,\bar\omega)\Bigr),\label{eq5-4}
\end{eqnarray}
\begin{eqnarray}
\tilde\nabla^\alpha \partial_\alpha\bar\omega -\frac{1}{6}\tilde R\bar\omega -\frac{\partial V}{\partial \bar\omega}-\frac{1}{9}\Lambda \kappa^2\bar\omega^3=0, \label{eq5-5}
\end{eqnarray}
\begin{eqnarray}
 D^\alpha D_\alpha\tilde\Phi-\frac{1}{6}\tilde R\tilde\Phi-\frac{\partial V}{\partial\tilde\Phi^*}=0,\cr \tilde\nabla^\nu F_{\mu\nu}=\frac{i}{2}\epsilon \Bigl(\tilde\Phi ( D_\mu\tilde\Phi)^* -\tilde\Phi^*  D_\mu\tilde\Phi\Bigr),\label{eq5-6}
\end{eqnarray}
with
\begin{eqnarray}
\tilde T_{\mu\nu}^{(A)}=F_{\mu\alpha}F_\nu^\alpha-\frac{1}{4}\tilde g_{\mu\nu}F_{\alpha\beta}F^{\alpha\beta},\label{eq5-7}
\end{eqnarray}
\begin{eqnarray}
\tilde T_{\mu\nu}^{(\tilde\Phi ,c)}&=&\Bigl(\tilde\nabla_\mu\partial_\nu\tilde\Phi\tilde\Phi^*-\tilde g_{\mu\nu}\tilde\nabla_\alpha\partial^\alpha\tilde\Phi\tilde\Phi^*\Bigr)\cr
&-&3\Bigl[ D_\mu\tilde\Phi( D_\nu\tilde\Phi)^*+( D_\mu\tilde\Phi)^* D_\nu\tilde\Phi \cr
&-&\tilde g_{\mu\nu} D_\alpha\tilde\Phi( D^\alpha\tilde\Phi)^*\Bigl]\label{eq5-8}
\end{eqnarray}
and
\begin{eqnarray}
T_{\mu\nu}^{(\bar\omega)}&=&\Bigl(\tilde\nabla_\mu\partial_\nu\bar\omega^2-\tilde g_{\mu\nu}\tilde\nabla_\alpha\partial^\alpha\bar\omega^2\Bigr)\cr
&-&6\Bigl(\partial_\mu\bar \omega\partial_\nu\bar \omega-\frac{1}{2}\tilde g_{\mu\nu}\partial_\alpha\bar \omega\partial^\alpha\bar\omega)\Bigl).\label{eq5-9}
\end{eqnarray}
Newton's constant reappears in the quadratic interaction term for the scalar field. The trace of Eq.(\ref{eq5-4}) is$\sim \kappa^2 \eta^2\lambda\bar\omega^2 X^2$.
It is not possible to isolate the equation for $\bar\omega$, as expected, We will return to this issue in section 5-B.
The resulting equations for $\bar\omega $, $\pmb{\psi}$, $\pmb{\gamma}$, $\pmb{\mu}$, X and P can then be solved numerically.
For the exterior we found Eq.(\ref{eq4-1})\footnote{{\tiny In order to avoid confusion with the dilaton $\omega$ of the interior spacetime, we will denote the exterior dilaton as $\pmb{\omega_E}$.}}
\begin{equation}
ds^2=\pmb{\omega_E}^2\Bigl[-dt^2+d\rho^2+e^{2\pmb{\tau_E}}dz^2+\rho^2 e^{-2\pmb{\gamma_E}}d\varphi^2\Bigr],\label{eq5-10}
\end{equation}
with exact solution Eq.(\ref{eq4-14}). The index $E$ refers to exterior.
The two line elements will be isometric at the boundary surface (core) $r=r_{c}(T)$ and $\rho=\rho_{c}(t)$, if
\begin{eqnarray}
\omega_E\rho e^{-\gamma}=\bar\omega r e^{-\tilde\psi},\qquad \omega_E e^{\tau}=\bar\omega e^{\tilde\psi+\tilde\mu} \cr
\quad \omega_E\sqrt{1-\dot\rho^2}dt = \bar\omega\sqrt{1-\dot r^2} dT.\label{eq5-11}
\end{eqnarray}
In addition, one needs matching conditions on the extrinsic curvature tensors of the boundary surfaces\cite{Anderson}.
\subsection{\label{sec:level 5-2}Connection with warped 5D conformal invariant model}
In the warped 5D counterpart model\cite{Slag1,Slag2}, we considered the spacetime
\begin{eqnarray}
ds^2 &=& {\cal W}(t,r,y)^2\Bigl[e^{2\gamma(t,r)-2\psi(t,r)}(-dt^2+ dr^2)\cr
&+&e^{2\psi(t,r)}dz^2+\frac{r^2}{e^{2\psi(t,r)}}d\varphi^2\Bigr]+ \Gamma(y)^2dy^2.\label{eq5-12}
\end{eqnarray}
The 4-dimensional brane is located at $y=0$. All standard model fields reside on the brane, while gravity can propagate into the bulk. ${\cal W}(t,r,y)$ is called the warp factor.
From the 5D Einstein equations it follows dat we can write ${\cal W}=W_1(t,r)W_2(y)$ and the exact solution becomes
\begin{eqnarray}
{\cal W}^2&=&\frac{e^{-\frac{1}{6} \Lambda_5(y- y_0)^2}}{\tau r}\Bigl(d_1 e^{(\sqrt{2\tau})t}-d_2e^{-(\sqrt{2\tau})t}\Bigr) \cr
&\times&\Bigl(d_3 e^{(\sqrt{2\tau})r}-d_4e^{-(\sqrt{2\tau})r}\Bigr).\label{eq5-13}
\end{eqnarray}
If we define
\begin{equation}
^{(5)}{g_{\mu\nu}}=W_1(t,r)^2 W_2(y)^2\tilde g_{\mu\nu}+n_\mu n_\nu \Gamma(y)^2,\label{eq5-14}
\end{equation}
then $W_1$ could be identified as dilaton field or "warp factor" coming from the bulk, while  $W_2$ is equals the well-know warp factor in the Randall-Sundrum brane world model\cite{Ran1,Ran2}.
The  metric $\tilde g_{\mu\nu}$ is equals our $g_{\mu\nu}=\omega^2\bar g_{\mu\nu}$ of Eq.(\ref{eq5-2}) of section 5-A
In the former study\cite{Slag2}, a numerical  solution was found in the coupled Einstein-scalar-gauge field model, and the trace anomaly could possibly be solve due to contributions from the bulk.
The solution for $W_1$ follows solely from the 5D Einstein equations with empty bulk ( eventually with a cosmological constant $\Lambda_5$).
The question is, can we identify $\omega$ with $W_1$?
After all, information about an extra dimension is visible as a curvature in a spacetime with one fewer dimension.
The $W_1$-field   need to be shifted to a complex contour to ensure that $W_1$ has the same unitary and positivity properties as the scalar field. One can easily check that $W_1$  has complex solutions.
When approaching smaller scales in the model and $W_1 \rightarrow0$, no singular behavior occurs due to the fact that the scalar field is present in the Lagrangian, which could be handled on the same footing as the dilaton field.
For later times, on cosmological scale,  $W_1$ behaves as a "standard" warp factor.
The $\Omega$-field is necessary as a conformal "gauge" in order to make a renormalizable model ( see Appendix C).
As 't Hooft stated\cite{thooft1}, "We get a renormalizable gauge if we decide to choose our conformal factor $\Omega$ in such a way that the amount of activity in a given spacetime volume is fixed or at least bounded".
It is then conjectured\cite{thooft1} that close to the Planck scale the conformal invariance is spontaneously broken due to the contributions of additional terms like  $ (R_{\mu\nu}R^{\mu\nu}-\frac{1}{3}R^2)$  and $\kappa^2\eta^2\lambda\bar\omega^2X^2$ in the action.
\section{\label{sec:level 6}Conclusions}
We investigated the coupled Einstein-scalar-gauge fields using  a conformally invariant Lagrangian.
We compare the exterior exact vacuum solution with the spacetime of an infinite line-mass and semi-infinite line-mass.
We  find a simple example of conformally equivalent (Ricci) flat spacetimes $\omega^2\bar g_{\mu\nu}$. This solution must then be matched on the non-vacuum solution of the cosmic string at the core.
In order to obtain tracelessness of the energy momentum tensor, additional constraints are necessary.
It is conjectured that a contribution from the bulk in the 5D counterpart model is needed to make the energy momentum tensor traceless.
The conformal symmetry, if exact, can be broken spontaneously. This means that we need additional field transformations on the vacuum spacetime.
Our analysis is purely classical and quantum corrections should be investigated.
In order to obtain an effective conformal invariant and finite  theory, many problems must be overcome, such as  unitarity violation and conformal anomalies.
In canonical gravity, quantum amplitudes are obtained by integrating the action over all components of the metric, so over $\omega$ and $\bar g_{\mu\nu}$  and imposing constraints on $\bar g_{\mu\nu}$ and matter fields (including fermionic fields)\cite{thooft3}.
In context with the 5D warped spacetime, the functional integration will be of the form
\begin{equation}
\int {\cal D}{^{\scriptscriptstyle{5}}\!{\cal W}}\int{\cal D}{^{\scriptscriptstyle{4}}\!\omega}\int{\cal D}\bar g_{\mu\nu}e^{iS}
\end{equation}
with $S$ the gauge fixing constraints.
We don't pretend we have solved the issue of the breaking of the conformal invariance (by anomalies).
We only describe in an example, how the Ricci-flat spacetimes could generate curvature  by a suitable dilaton field $\omega$  and gauge freedom $\Omega$.
\appendix
\section{\label{sec:level A}The Semi-Infinite Line Mass and Accelerated Observers}
For the spacetime
\begin{equation}
ds^2=e^{-2\psi}\Bigl[e^{2\gamma}(d\rho^2+dz^2)+\rho^2d\varphi^2\Bigr]-e^{2\psi}(dt+Wd\varphi)^2,\label{eqA-1}
\end{equation}
where $\psi, \gamma$ and $ W$ are functions of $\rho$ and $z$, one can generate many solutions. We refer to Islam\cite{Islam} for a nice overview.
The solution of Eq.(\ref{eq2-3}) is an example of a solution of Eq.(\ref{eqA-1}) (with $W=0$).
This  Ricci-flat solution has some interesting properties. The constant $\alpha$ can be transformed away, except for $c_1=1$. Then a genuine constant besides $c_1$ arises. The Kretschmann-scalar K for $\rho=0$
becomes
\begin{equation}
K=\frac{12 c_1^2(2c_1-1)^2[2(z-z_1)]^{4c_1}}{\alpha^2(z-z_1)^2}.\label{eqA-2}
\end{equation}
Solutions for $c_1>1$ are physically unlikely, because the proper distance from $z_0>z_1$ to infinity becomes  finite for $c_1>1$. So one cannot reach infinity along this direction.
If one  performs the transformation
\begin{equation}
\rho=z'\rho ',\quad 2\epsilon(z-z_1)=z'^2-\rho'^2\label{eqA-3}
\end{equation}
on Eq.(\ref{eq2-3}), one obtains
\begin{eqnarray}
ds^2=-(z')^{4c_1}dt^2&+&(z')^{8c_1^2-4c_1}(\rho'^2+z'^2)^{1-4c_1^2}(dz'^2+d\rho'^2)\cr
&+&\rho'^2(z')^{2-4c_1}d\varphi^2,\label{eqA-4}
\end{eqnarray}
which maps the spacetime onto the half space $z'>0$ and $\rho'\geq 0$. For $c_1=\frac{1}{2}$ the metric becomes
\begin{equation}
ds^2=-z'^2dt^2+dz'^2+d\rho'^2+\rho'^2 d\varphi^2,\label{eqA-5}
\end{equation}
which represents a uniformly accelerating metric\cite{Rindler}.
Points on the $z'$-axis have constant acceleration relative to Minkowski spacetime.
For $c_1=0$ and $c_1=\frac{1}{2}$ the spacetime is flat. So we have a strange behavior of this peculiar spacetime. If we increase the mass density from $c_1=0$ to $c_1=\frac{1}{2}$, we obtain again a flat spacetime.
The Rindler-transformation
\begin{eqnarray}
z=\frac{1}{2}(z'^2-t'^2-\rho'^2)&,&\qquad \rho=\rho'(\sqrt{z'^2-t'^2})
\cr &t&=arctanh(\frac{t'}{z'}),\label{eqA-6}
\end{eqnarray}
brings the spacetime again to a half-Minkowski, $ds^2=dz'^2+d\rho'^2+\rho'^2d\varphi^2-dt'^2$, with $z'^2\geq t'^2$.
\begin{figure}[h]
\centerline{\includegraphics[scale=.5]{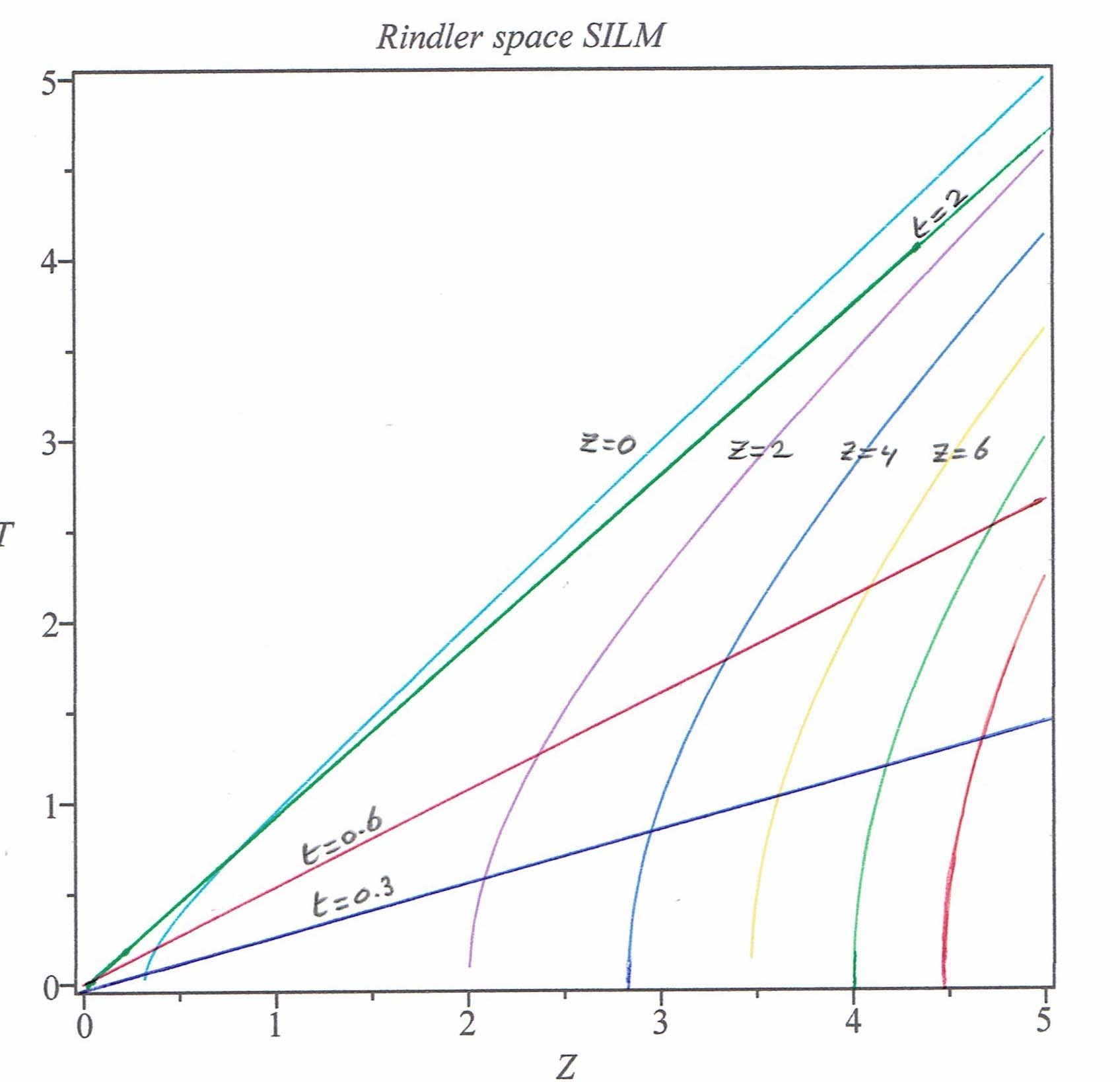}}
\caption{{\bf The Rindler diagram for the SILM}}
\label{fig:6}
\end{figure}
The SILM can be analysed in Rindler-coordinates.
A Rindler-observer at rest in  Rindler-coordinates,  has a constant proper acceleration.  The Rindler-observer whose proper time is equals the coordinate time, is the one who has proper acceleration 1.
Rindler-observers close to the horizon $Z=0$ have a greater proper acceleration. All Rindler-observers are at rest at time $T=0$ in the inertial frame and at this time a Rindler-observer with proper acceleration $g$ will be at position $X=\frac{1}{g}$, which is also the distance from the Rindler-horizon in Rindler-coordinates.
In figure 6 we plotted the Rindler-diagram for the expressions following from the transformation of Eq.(\ref{eqA-6})
\begin{eqnarray}
z'^2-t'^2=z\pm\sqrt{z^2+\rho^2}&,&\qquad t'=z'\tanh(t) \cr
t'^2&=&\frac{\rho^2\sinh^2(t)}{\rho'^2}.\label{eqA-7}
\end{eqnarray}
If we clue together the two solutions $\epsilon=\pm 1$ of Eq.(\ref{eq2-3}), we obtain the infinite line mass (ILM).
A suitable form will be
\begin{equation}
ds^2=-\alpha \rho^{4c_1}dt^2+\frac{1}{\alpha}\rho^{8c_1^2-4c_1}(d\rho^2+dz^2)+\frac{1}{\alpha}\rho^{2-4c_1}d\varphi^2.\label{eqA-8}
\end{equation}
For an ILM  one cannot ignore a finiteness of the line mass. One needs a description of the interior, expressed by a second constant besides the mass density. The relation with the cosmic strings will then be clarified.
This metric is also known as the Levi-Civita spacetime (the Levi-Civita spacetime plays a fundamental role in constructing conformal equivalent spacetimes).
This metric can be transformed to
\begin{equation}
ds^2=-\rho^{4c_1}dt^2+\rho^{8c_1^2-4c_1}(d\rho^2+dz^2)+\frac{1}{\alpha}\rho^{2-4c_1}d\varphi^2,\label{eqA-9}
\end{equation}
by redefinition of $ \rho=a \rho', z=b z'$ and $t=c t'$ for some constants $a, b,$ and $c$. We have again  two constants, i. e.,  the mass per unit length $c_1$ and $\alpha$ determined by the internal composition of the cylinder. If we try to scale away $\alpha$, we obtain an angle deficit. In the spherically symmetric case where the general relativistic solutions contains only one constant, conservation law than ensures that this constant is conserved and the spacetime is then necessarily static. In the cylindrical case this is not the case, because energy may flow to and from infinity in the axial direction.
The Kretschmann-scalar is
\begin{equation}
K=\frac{64c_1^2(4c_1^2-2c_1+1)(2c_1-1_)^2}{\rho^{4(4c_1^2-2c_1+1)}}.\label{eqA-10}
\end{equation}
It is infinite for $\rho =0$, except for $c_1=0,\frac{1}{2}$.
If we calculate the proper distance $\int\sqrt{g_{\rho\rho}}d\rho$ and evaluate K at unit proper distance, we obtain $K=\frac{64c_1^2(2c_1-1)^2}{4c_2-2c_1+1)^3}$. In figure 7 we plotted the Kretschmann scalar.
\begin{figure}[h]
\centerline{\includegraphics[scale=.1]{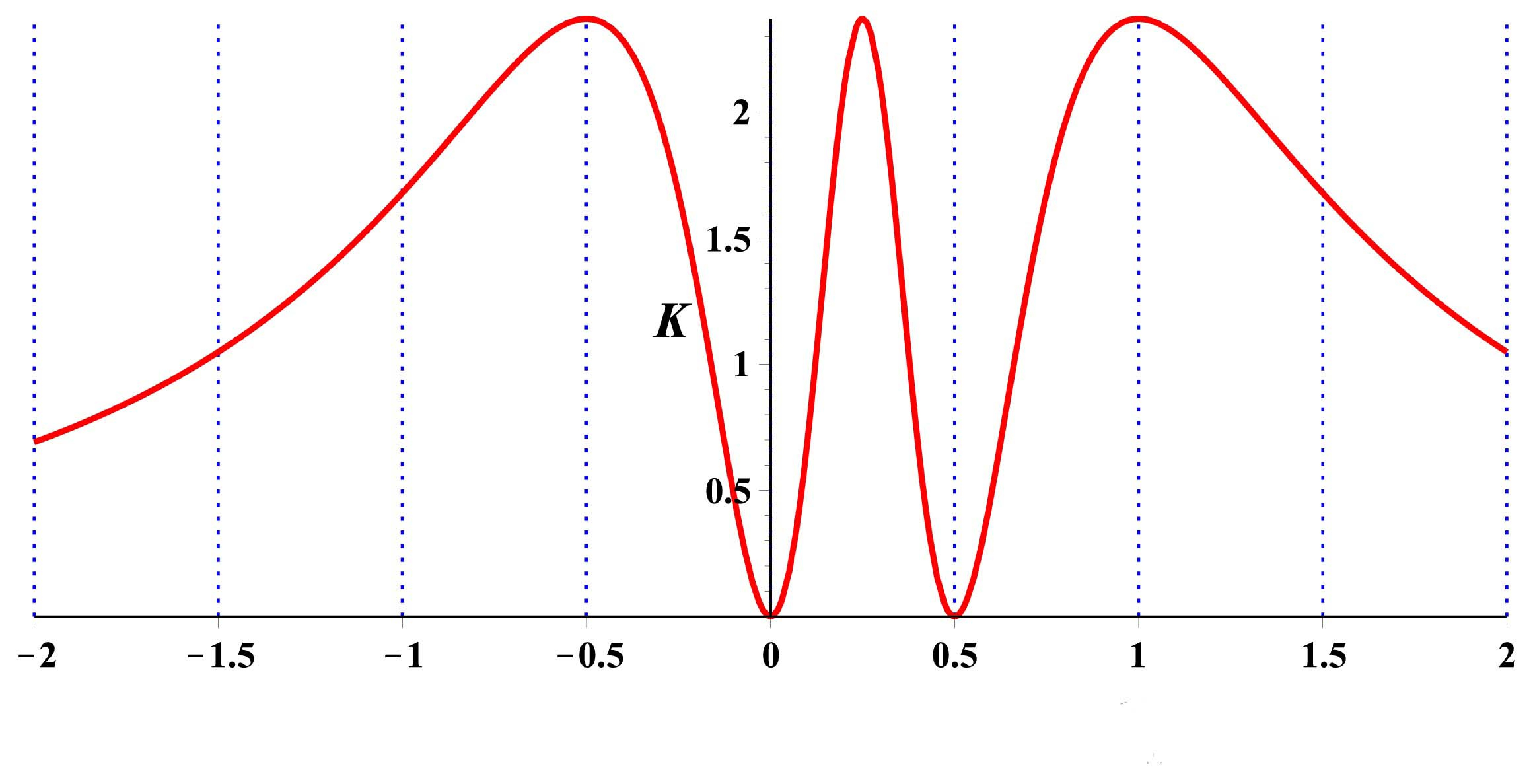}}
\caption{{\bf The Kretschmann-scalar plotted as function  of $c_1$}}
\label{fig7}
\end{figure}
It is remarkable that we obtain for $c_1=0$ the feature of a cosmic string on a whole other level.
However, the angle deficit is   determined by the symmetry breaking scale and the gauge-to scalar mass ratio. So without the detailed knowledge of the matter distribution  we expect an angle deficit for the the ILM. The constant $\alpha$ will then also  contain the mass per unit length determined by the scalar gauge fields. It is inevitable to ignore the interior of the cylinder.
The physical  behavior can change abruptly by a change in the two parameters.
For $c_1 <\frac{1}{2}$ one can match the solution on an interior solution. The dimension of the core of the string will then enter the model.
It is conjectured\cite{Herr} that the $c_1=\frac{1}{2}$ solutions describes a kind of planar mass distribution. However, the Riemann tensor vanishes. So the question remains: why does a cylinder with positive energy density and pressure produces vanishing curvature. We shall see that without the cosmic string interpretation, one cannot explain this remarkable feature.
The dependance of the exterior solution on two parameters has a strong bearing on the existence of gravitational waves, if we make the transformation $t\rightarrow iz, z\rightarrow it$.
\section{\label{sec:level B}Prelude of the ILM Inconsistency}
For a stringlike object expects that at a finite radial distance the field variables are  properly matched on the  vacuum solution. Alternatively, this can occur  asymptotically at radial infinity, i.e., in the case of the ILM Eq.(\ref{eqA-8}) with $c_1=0$ and $\frac{1}{2}$. This must hold even in the wire-approximation, where the radius $\rho_0$ of the string is of the order $10^{-30}$ cm.

An attempt to find an interior solution to the Levi-Civita exterior solution was done by Gott\cite{Gott} and Hiscock\cite{Hiscock}(GH). They consider the stress-energy tensor $T_t^t=T_z^z=\mu, T_\rho^\rho=T_\varphi^\varphi=0$.
The interior solution they found is
\begin{equation}
ds^2=-dt^2+dr^2+dz^2+a^2\sin^2(\frac{\rho}{a})d\varphi^2.\label{eqB-1}
\end{equation}
So the solution does not contain any trace of the the gauged scalar field. So it not a surprise that the matching between the interior and exterior solutions will be  inconsistent\cite{Ray}.
The GH-solution can be interpreted as an approximation.
In the special case where the mass of the gauge field is equals the mass of the scalar field (Bogomol'nyi bound), one obtains from $T_\rho^\rho=T_\varphi^\varphi=0$  the field equations
\begin{eqnarray}
\partial_\rho P=\frac{1}{2}\sqrt{g_{\varphi\varphi}}e^2(X^2-\eta^2),\qquad \partial_\rho X=\frac{XP}{\sqrt{g_{\varphi\varphi}}},\\
\partial_\rho\sqrt{g_{\varphi\varphi}}=-4\pi P(X^2-\eta^2)+1-4\pi \eta^2.\label{eqB-2}
\end{eqnarray}

One obtains the GH-solution if one neglects terms of order $X^2$
P and X can also be  expressed in trigonometrical functions. As boundary conditions one needs
\begin{eqnarray}
X(\rho)\rightarrow \rho ( \rho \rightarrow 0), \qquad P(0)=1, \cr
\sqrt{g_{\varphi\varphi}}\rightarrow\rho ( \rho\rightarrow 0), \quad X(\rho_0)=1,\quad P(\rho_0)=0.\label{eqB-3}
\end{eqnarray}
For the exterior solution, one then obtains $g_{\varphi\varphi}=b^2\rho^2$, which is the Minkowski metric minus a wedge. From the Darmois-Lichnerowicz matching conditions at the boundary,  one  easily obtains  $ b=(1-4\mu)$, where $\mu$ represents the linear energy density determined by $ \eta$ and $\rho_0$. The angle deficit is  $\triangle \varphi \approx 8\pi\mu$.
In general, there doesn't exist a simple relation between $\triangle \varphi$ and $\mu$\cite{Futgar}. The "thin"-string approximation is not applicable without severe restrictions.
One can say that the this approximation doesn't accurately reflect the properties of finite-thickness GR cosmic string in the zero-thickness limit.
The one-parameter Minkowski metric minus a wedge could be a thin-string approximation in  GR and can be considered as the gravitational field of an infinite thin wire with distributional stress energy, if the radial stress is negligible compared with the energy density\cite{Israel}. This is impossible for cosmic string solutions. Geroch and Tasschen\cite{Gertas} consider "regular" metrics, where the curvature tensor make sense as a distribution and demonstrate that the metric of an infinite thin string cannot be regular and cannot assign a distributional stress-energy tensor to Minkowski minus a wedge.
They conclude also, that the approximate relation $\triangle \varphi \approx 8\pi\mu$ is  not valid without severe  (but unrealistic) restrictions on the stress-energy tensor.
This holds also for the general self-gravitating cosmic string of section 2. The field equations depend on the parameters $\alpha =\frac{e^2}{\lambda}$ and $\eta$. We should like to find a one-parameter subfamily which has a zero-thickness limit. One can define a "scaling" family that is a subfamily defined by the condition that $\eta $ and $\alpha$ are constant. This condition gives us a one-parameter curve in parameter space, so a one-parameter family of cosmic strings with parameter $\lambda$. If one defines a scaling transformation to be a change in $\lambda$ and e which leaves $\alpha$ fixed, then the field equations are invariant under this re-scaling and are physically equivalent. The zero-thickness limit can then be achieved by $\lambda \rightarrow \infty$.
Since the fields change continuously with $\rho$, there is no radius where the core of the cosmic strings abruptly ends. As  suitable "effective" radius of the core can be the coherence length $\zeta=\frac{1}{\eta\sqrt{\lambda}}$, which is just the scale transformation  $\rho\rightarrow\frac{\rho}{\eta\sqrt{\lambda}}$. In section 2 we  used this transformation to  derive the two-parameter field equations.
So the scaling transformation changes the size of the string.
In the limit $\lambda\rightarrow\infty$, the effective radius approaches zero, i.e., a zero thickness in a scaling family of cosmic strings. However, the energy blows up in this limit.
The expression for the mass per unit length, $\sigma = 2\pi\int\sqrt{g_{\varphi\varphi}}T_{tt} d\rho$, is also invariant under the scaling transformation and so is the angle deficit . In the weak-field limit, when $\eta\rightarrow 0$, the angle deficit approaches $8\pi\sigma$. The correction term  is also scale invariant.
The Bogomol'nyi bound is not applicable if we consider $(t,\rho)$-dependency and the features of the string as described here are unlikely.
We shall see that  the $(t,\rho)$-dependency is mandatory in order to describe the correct matching conditions in the full model. This is not surprising, because the stationary situation can be transformed to the $(t,\rho)$-dependency and gravitational waves enter the scene.
\section{\label{sec:level C}Conformal Invariance}
Conformal transformations occur in many physical contexts, specially in GRT. Conformal invariance  plays a role in the notion of asymptotic flatness and causal structure of isolated systems when they radiate\cite{Wald}.
If one considers a field ${\cal F}$ on a metric $g_{\mu\nu}$ , one says that $\Omega^s {\cal F}$ is conformally invariant  with metric $\Omega^2 g_{\mu\nu}$ for all conformal factors $\Omega^2$.
s is called the conformal weight of the matter field. Mathematically, a conformal transformation will preserve angles, but it changes the magnitudes of lengths of vectors.
The Maxwell equations are conformally invariant, but the (massless) Klein-Gordon equation $\nabla^2\Phi=0$ is not. The conformal invariant Klein-Gordon equation (in n-dimensions) can be written as
\begin{equation}
\Bigl(\nabla^2-\frac{n-2}{4(n-1)}R\Bigr)\Phi=0,\label{eqC-1}
\end{equation}
Einstein's equations are not conformal invariant.
Under a conformal transformation  $g_{\mu\nu}\rightarrow\Omega^2 g_{\mu\nu}$, $G_{\mu\nu}$ transforms  as\cite{Wald}
\begin{eqnarray}
G_{\mu\nu}\rightarrow G_{\mu\nu} &+&\frac{2}{\pmb{\Omega}^2}\Bigl[2\nabla_\mu\pmb{\Omega}\nabla_\nu\pmb{\Omega}-\pmb{\Omega}\nabla_\mu\nabla_\nu\pmb{\Omega}\cr
&-&\frac{1}{2}  g_{\mu\nu}(\nabla^\alpha\pmb{\Omega}\nabla_\alpha\pmb{\Omega}-2\pmb{\Omega}\nabla^\alpha\nabla_\alpha\pmb{\Omega})\Bigr]\label{eqC-2}
\end{eqnarray}
and the Ricci scalar as
\begin{equation}
R\rightarrow\frac{1}{\pmb{\Omega}^2}\Bigl(R-\frac{6}{\pmb{\Omega}}\nabla^\mu\nabla_\mu\pmb{\Omega}\Bigr).\label{eqC-3}
\end{equation}
Further, $\sqrt{-g}$ transforms as $\sqrt{-g}\rightarrow\Omega^4\sqrt{-g}$.
The conformal invariant Einstein-Hilbert action becomes
\begin{equation}
{\cal S}=\int dx\sqrt{-g}\Bigl[\Omega^2R-2\Omega^4\Lambda +4\frac{n-1}{n-2} g^{\mu\nu}\partial_\mu\Omega\partial_\nu\Omega\Bigr].\label{eqC-4}
\end{equation}
By varying the action with respect to  $g_{\mu\nu}$ and $\Omega$, one then obtains the field equation of  section 4-A.
If we write $g_{\mu\nu}=\Omega^2 \tilde g_{\mu\nu}$, then on a Lorentz metric, $g_{\mu\nu}$ and $\tilde g_{\mu\nu}$ will have the same causal structure. On a non-Lorentzian spacetimes this is not necessarily true.
We observe from   Eq.(\ref{eqC-3}) that the  vacuum  $R=0$ is conformal invariant, if $\nabla^\mu\partial_\mu\pmb{\Omega}=0$.
In the case of Eq.(\ref{eq4-1}), we obtain
\begin{equation}
\partial_{tt}\pmb{\Omega}-\partial_{\rho\rho}\pmb{\Omega} -\frac{1}{\rho}\partial_\rho\pmb{\Omega}+\partial_t\pmb{\Omega}\partial_t\tau
-\partial_\rho\pmb{\Omega}\partial_\rho\tau -\partial_t\pmb{\Omega}\partial_t\gamma
+\partial_\rho\pmb{\Omega}\partial_\rho\gamma =0,\label{eqC-5}
\end{equation}
which can be solved for $\tau$ and $\gamma$ satisfying Eq.(\ref{eq4-2}) and Eq.(\ref{eq4-3}) .  An example is the  solution
\begin{equation}
\pmb{\Omega}={\cal G}(r+t)+{\cal H}(r-t)e^{-\frac{1}{2}{\cal F}(r+t)-t}\label{eqC-6}
\end{equation}
with ${\cal G}$ and ${\cal H}$ arbitrary functions of $(r+t)$ and $(r-t)$ respectively.
However, a better way is to proceed with the method  in section 4-A, where the metric field equations and the dilaton field equations are solved  simultaneously.

\bibliography{apssamp}

\begin{thebibliography}{999}
\bibitem{Step}
Stephani, H., Kramer, D., Maccallum, M. and Herlt, E (2009) {\em Exact Solutions of Einstein's Field Equations} , Cambridge University Press, Cambridge, UK.
\bibitem{Islam}
Islam, J. N., (1985) Rotating Fields in General Relativity, Cambridge University Press, Cambridge, UK.
\bibitem{Fel}
Felsager, B. (1987) {\em Geometry, particles and fields}, Odense univ.press: Odense, {\bf 1987}.
\bibitem{Nielsen}
Nielsen H B and Olesen P, (1973) {\em Nucl.Phys.} B {\bf 61}, 45
\bibitem{Garf}
Garfinkle, D.; (1985) General relativistic strings.  {\em Phys. Rev. D}, {\bf 32}, 1323.
\bibitem{Shir}
Shiromizu, T.,  Maeda, K. and Sasaki, M. (2000) {\em Phys. Rev. D} {\bf 62}, 024012.
\bibitem{Slag1}
Slagter, R. J. and Pan, S.(2016) New fate of a warped 5D FLRW model with a U(1) scalar gauge field. {\em Found. of Phys.} {\bf 46}, 1075.
\bibitem{Slag2}
Slagter, R. J. (2017)   submitted to {\em Found. of Phys.}, ArXiv:: gr-qc/1711.08193
\bibitem{ER}
Einstein, A. and Rosen, N. (1937) {\em J. Franklin Inst. } {\bf 223}, 43.
\bibitem{thooft1}
’t Hooft, G., (2015), gr-qc/151104427v1
\bibitem{thooft2}
’t Hooft, G., (1993), gr-qc/9310026
\bibitem{thooft3}
’t Hooft, G., (2010), gr-qc/10090669v2
\bibitem{thooft4}
’t Hooft, G., (2010), gr-qc/10110061v1
\bibitem{thooft5}
’t Hooft, G., (2011), {\it Found. of Phys.} {\bf 41}, 1829.
\bibitem{thooft0}
’t Hooft, G. and Veltman, M. (1974), {\em Ann. Inst. Henri Poincare} {\bf 20}, 69.
\bibitem{Parker}
Parker, L. E. and Toms, D. J. (2009) Quantum Field Theory in Curved Spacetime, Cambridge University Press, Cambridge, UK.
\bibitem{Mann1}
Mannheim, P. D., (2017), hep-th/161008907v2
\bibitem{Mann2}
Mannheim, P. D., (2017),{\em  J. of Phys. G: Nucl. and Part. Phys.} {\bf 44} hep-th/161008907v2
\bibitem{Stachel}
Stachel, J. J., (1986)  {\em J. Math. Phys.} {\bf 7}, 1321.
\bibitem{Marder}
Marder, L. (1958) {\em Proc. R. Soc. A} {\bf 244}, 524.
\bibitem{Bondi}
Bondi, H. (1989)  {\em Proc. R. Soc. Lond. A}  {\bf 427}, 259
\bibitem{Thorne}
Thorne, K. S. (1965) {\em Phys. Rev. B} {\bf 138}, 251.
\bibitem{Vil}
Vilenkin, A., Shellard, E. P. S. , (1994) {\em Cosmic strings and other topological defects}; P. V. Lanshoff, et  al, Eds. Cambridge univiversity press: Cambrigde  UK.
\bibitem{Bonner1}
Bonner, W. B. (1979) {\em J. Phys.A: Math. Gen} {\bf 13}, 2121
\bibitem{Bonner2}
Bonner, W. B. (1991) {\em Gen. Rel. Gravit.} {\bf 24}, 551.
\bibitem{Bonner3}
Bonner, W. B., Griffiths, J. B. and MacCallum, M. A. H. (1994) {\em Gen. Rel. Gravit.} {\bf 26}, 687.
\bibitem{Bonner4}
Bonner, W. B. and Martins, M. A. P. (1991) {\em Class. Quant. Grav.} {\bf 8}, 727.
\bibitem{Unruh}
Unruh, W. G. (1976) {\it Phys. Rev. D} {\bf 14}, 870.
\bibitem{Wald}
Wald, R. M.  (1984) {\em General Relativity}, University of Chicago Press, Chicago,  USA.
\bibitem{Anderson}
Anderson, M.R. (2003) {\em The Mathematical Theory of Cosmic Strings}. IoP publishing, Bistol, UK.
\bibitem{Ran1}
Randall, L. and Sundrum, R. (1999) {\it Phys. Rev. Lett.}, {\bf 83},3370.
\bibitem{Ran2}
Randall, L. and Sundrum, R. (1999) {\it Phys. Rev. Lett.} {\bf 83}, 4690.
\bibitem {Rindler}
Rindler, W. (2006) {\em Relativity}. Oxford University Press, Oxford, UK.
\bibitem{Herr}
Herrera, L., Santos, N. O., Teixeira, A. F. F. and Wang, A. Z. , (2001) {\em Class. Quant. Grav.} {\bf 18}, 3847.
\bibitem{Gott}
Gott, J. R. (1985) {\em Astrophys. J.} {\bf 288}, 422.
\bibitem{Hiscock}
Hiscock, W. A., (1985) {\em Phys. Rev. D} {\bf 31}, 3288.
\bibitem{Ray}
Raychaudhjuri, A. K. (1990) {\it Phus. Rev D} {\bf 41}, 3041.
\bibitem{Futgar}
Futamase, T. and Garfinkle, D. (1988) {\em Phys. Rev. D.} {\bf 37}, 2086.
\bibitem{Israel}
Israel, W. (1977) {\em Phys. Rev. D.} {\bf 15}, 935.
\bibitem{Gertas}
Geroch, R. and Traschen, J. (1987) {\em Phys. Rev. D.} {\bf 36}, 1017.
\bibitem{strom}
Strominger, A. (2001) {\em Journal of High Energy Physics.}  {\bf 1}, 34.
\bibitem{Unruh}
Unruh, W. B. (1967) {\em Phys. Rev. D.} {\bf 14}, 870.

\end{thebibliography}
\section*{References}

\end{document}